\documentstyle[aps,amstex,amssymb,preprint,epsf]{revtex}
\begin{document}                                                       

\draft 

\preprint{UASLP--IF--00--03}
                                            
\title {Control of tunneling by adapted signals}

\author{B. I.\ Ivlev}

\address{Department of Physics and Astronomy\\
University of South Carolina, Columbia, SC 29208\\
and\\
Instituto de F\'{\i}sica, Universidad Aut\'onoma de San Luis Potos\'{\i}\\
San Luis Potos\'{\i}, S. L. P. 78000 Mexico}

\maketitle

\begin{abstract}
Process of quantum tunneling of particles in various physical systems can be effectively controlled even by a weak and slow
varying in time electromagnetic signal if to adapt specially its shape to a particular system. During an under-barrier 
motion of a particle such signal provides a ``coherent'' assistance of tunneling by the multi-quanta absorption resulting 
in a strong enhancement of the tunneling probability. The semiclassical approach based on trajectories in the complex time
is developed for tunneling in a non-stationary field. Enhancement of tunneling occurs when a singularity of the signal 
coincides in position at the complex time plane with a singularity of the classical Newtonian trajectory of the particle. 
The developed theory is also applicable to the over-barrier reflection of particles and to reflection of classical waves 
(electromagnetic, hydrodynamic, etc.) from a spatially-smooth medium.
\end{abstract} \vskip 1.0cm


\narrowtext
\section{INTRODUCTION}
Control of quantum systems by tuned external signals is an actively developed field at present, see for example 
\cite{WARREN} and references therein. Excitation of molecules, when one should excite only particular chemical bonds 
\cite{SHI,JUDSON,KOHLER}, formation of programmable atomic wave packets \cite{SCHUM}, a control of electron states in 
heterostructures \cite{KRAUSE}, and a control of photo-current in semiconductors \cite{ATAN} are typical examples of control
by laser pulses. A control of quantum tunneling by electromagnetic signals is also a matter of interest, since tunneling 
is a part of many physical processes and of some chemical reactions. Modern facilities enable to tailor fast signals 
required for this purpose \cite{WEINER,EFIMOV}.
 
Let us focus on main aspects of tunneling under a non-stationary perturbation. The potential barrier $V(x)$, extended 
over the distance $a$, sets two typical energy scales: the barrier hight $V_0$ and $\hbar\omega\sim\hbar\sqrt{V_{0}/ma^{2}}$, where 
$\omega$ can be associated with an oscillation frequency in the overturned potential. For semiclassical barriers the two 
energy scales are well separated $\hbar\omega\ll V_0$ and in absence of a non-stationary field the probability of tunneling 
through the barrier can be estimated with the exponential accuracy as $W\sim\exp(-V_{0}/\hbar\omega)$. In presence of the 
periodic signal ${\cal E}_{\Omega}\cos\Omega t$ a particle can absorb the quantum $\Omega$ with the probability 
$(a{\cal E}_{\Omega}/V_{0})^2$ and tunnel in the more transparent part of the barrier with the probability 
$\exp(-(V_{0}-\Omega)/\hbar\omega)$. The total tunneling rate can be written as
\begin{align}
\label{1}
W&\sim\exp\left(-\frac{V_{0}}{\hbar\omega}\right)+
\left(\frac{a{\cal E}_{\Omega}}{V_{0}}\right)^{2}\exp\left(-\frac{V_{0}-\hbar\Omega}{\hbar\omega}\right)\nonumber\\
&=\exp\left(-\frac{V_{0}}{\hbar\omega}\right)\left(1+\left(\frac{a{\cal E}_{\Omega}}{V_{0}}\right)^{2}{\rm e}^{\Omega/\omega}\right)
\end{align}
Eq.~\ref{1} is approximate since in quantum mechanics one should multiply amplitudes but not probabilities, nevertheless
the form (\ref{1}) accounts necessary physical feature. Suppose a tunneling particle is acted by some electric field 
${\cal E}(t)$ (periodic or pulse-type) and the typical time scale of this signal is $\theta$. Than one can distinguish two 
different physical situations: (i) {\it hard signal}, fast varying field with $\theta\sim\hbar/V_{0}$ and (ii) {\it soft signal}, 
slow varying field with $\theta\sim\omega^{-1}$. Tunneling can be governed easily by a hard signal even when its amplitude is 
less than the static barrier field $V_{0}/a$, since the probability of quantum absorption $(a{\cal E}/V_{0})^2$ competes, 
according to Eq.~\ref{1}, with the small tunneling rate (in this case $\Omega\sim\theta^{-1}\sim V_{0}/\hbar$). It is also 
obvious that a soft signal of the high amplitude $V_{0}/a$ (static field of the barrier) is able to govern tunneling. Can a
soft signal with the amplitude much smaller than the static field of a potential barrier control effectively a tunneling 
process? Suppose a soft signal has the following shape
\begin{equation}
\label{2}
{\cal E}(t) =\frac{{\cal E}}{(1+t^{2}/\theta^{2})^{n}}
\end{equation}
Its Fourier harmonics ${\cal E}_{\Omega}\sim(\Omega\theta)^{n-1}{\cal E}\theta\exp(-\Omega\theta)$ should be inserted into Eq.~\ref{1}.
As follows from Eq.~\ref{1}, when the signal width $\theta$ is less than $1/2\omega$, the quadratic in ${\cal E}$ correction
to the static probability diverges with increase of $\Omega$. It means the perturbation theory with respect to a weak 
non-stationary signal to break down for sufficiently short pulses. Note, the pulse still remains soft. This is an 
indication of efficiency of soft signals. Breaking of the perturbation theory means a significance of multi-quantum 
processes and a principal question is that what theory has to be used in this case.

A review of some aspects of tunneling in complex systems, including the instanton approach, was done in the book 
\cite{LEGGETT}, see also Ref.~\cite{HANGGI}. Recent achievements in the semiclassical theory under stationary conditions are
presented in Refs.~\cite{KESHA,MAITRA,ANKER}. As it has been argued in Refs.~\cite{MELN,MELN1,LEGGETT}, the 
semiclassical method of complex trajectories is applicable also to a non-stationary case, when a signal is periodic in 
time. Nevertheless, despite a number of publications, use of semiclassical theory for tunneling in a non-stationary field 
remains non-obvious. What happens in general case, in particular, for a short pulse like one given by the relation 
(\ref{2})? The goal of this work is to show that the semiclassical theory based on the concept of the complex time is an 
appropriate description of tunneling under action of a soft pulse of any shape. For the particular case of a triangular 
barrier the tunneling rate in presence of a non-stationary (soft) field is found to be determined, in the main exponential 
approximation, by the classical action $S(x,t)$ satisfying the Hamilton-Jacobi equation. In this approximation the wave 
function is proportional to $\exp(iS(x,t))$ (Below Planck's constant is unity). The first correction (preexponential) to this 
classical result and the second one were found explicitly and shown to be small comparing to the main contribution, what 
is typical for semiclassical approximation. The tunneling rate is found as a function of time, it tends to its static 
value at $t\rightarrow\pm\infty$, when ${\cal E}(t)=0$, and reaches the maximum at some moment of time. This maximum value is 
given, with an exponential accuracy, by an extreme value of the classical action, which is determined, according to 
classical mechanics, by means of classical trajectories. The classical trajectory obeys Newton's equation 
$m\partial^{2}x/\partial t^{2}+V^{\prime}(x)={\cal E}(t)$ in the complex time, since in the real time there is no classical 
under-barrier path.

So, the method of classical trajectories in the complex time can be used, when the full time dependence of the tunneling 
rate is not required, but only its maximum value with an exponential accuracy is a matter of interest. Note, despite the 
classical action depends functionally on trajectories defined in the complex time plane, the argument $t$ of the action 
$S(x,t)$ is considered to be always real. The complex time has no physical meaning, it is only a possible way to 
parameterize a solution of the Hamilton-Jacobi equation. Nevertheless, this way is very useful. For a weak non-stationary 
field ${\cal E}(t)$ the classical trajectory $x_{0}(t)$ satisfies the unperturbed equation 
$(m/2)(\partial x_{0}/\partial t)^{2}+V(x_{0})=E$, where $E$ is the particle energy, which can identified with an incident energy of
particle flux on the barrier $V(x)$. The classical trajectory $x_{0}(t)$, as a function of the complex $t$, has the 
singularity at $t=t_{s}(E)$ \cite{MELN,MELN1} and the external signal (\ref{2}) has the singularity at $t=i\theta$. As shown in 
this paper, when the two singularities coincide
\begin{equation}
\label{3}
{\rm Im}\hspace{0.1cm}t_{s}(E)=\theta
\end{equation}
the effect of an external signal on tunneling enhances. Under the condition (\ref{3}) the perturbation theory breaks down 
at essentially weaker non-stationary signal comparing to a general case. The parameter ${\rm Im}\hspace{0.1cm}t_{s}(E)$ depends 
on properties of the static barrier and the particle energy \cite{MELN,MELN1}, but $\theta$ is a characteristic of a 
non-stationary signal. The physical meaning of the condition (\ref{3}) is not straightforward since a quantum mechanical 
process is described by a product of amplitudes but not of probabilities. The condition (\ref{3}) rather corresponds to a 
coherent cooperation of tunneling and quanta absorption, in other words, it is some ``resonance'' condition between 
motion of system and external signal. Eq.~\ref{3} determines some remarkable threshold energy $E_{T}$. As shown below, when 
the particle energy is big $E>E_{T}$ the tunneling process is moderately violated by the signal (\ref{2}) (of course, if the 
signal is less, than the static field of the barrier $V/a$); for lower energy $E<E_{T}$ the process of the barrier 
transition is strongly stimulated even by a relatively small signal. This opens a possibility to manipulate effectively a 
tunneling process by a specially adapted (according to Eq.~\ref{3}) electromagnetic signal of a small amplitude.

In Sections II-VII the tunneling probability as a function of time is calculated for the triangular barrier. In Section 
VIII the method of complex trajectories is described. In Sections IX-X this method is applied to the triangular barrier to 
compare with the results obtained by the direct solution of the Hamilton-Jacobi equation. In Sections XI-XIV the method of
complex trajectories is applied to a barrier given by an analytical function $V(x)$, when there is no simple solution of 
the Hamilton-Jacobi equation.
\section{TRIANGULAR BARRIER}
In this Section we consider decay of the metastable state in the potential
\begin{equation}
\label{4}
V(x)=V-{\cal E}_{0}\mid x\mid -\sqrt{\frac{2(V-E)}{m}}\hspace{0.1cm}\delta(x)
\end{equation}
under action of a non-stationary electric field ${\cal E}(t)$. In the limit ${\cal E}_{0}\rightarrow 0$ the energy $E$ corresponds
to the bound state in the $\delta$-function potential well. The symmetric wave function ($\psi(x,t)=\psi(-x,t)$) can be
written down in the form
\begin{equation}
\label{4a} 
\psi(x,t)=\exp(iS(x,t)+i\sigma(x,t))
\end{equation}
where the classical action $S$ obeys the Hamilton-Jacobi equation
at $x>0$
\begin{equation}
\label{5}
\frac{\partial S}{\partial t}+\frac{1}{2m}\left(\frac{\partial S}{\partial x}\right)^{2}+V -{\cal E}_{0}x-x{\cal E}(t)=0
\end{equation}
with the boundary condition
\begin{equation}
\label{6}
\left(\frac{\partial S(x,t)}{\partial x}\right)_{x=0}=i\sqrt{2m(V-E)}
\end{equation}
At $x=0$ one can impose the condition $S(0,t)=-Et$.  
The equation for $\sigma$ has the form 
\begin{equation}
\label{7}
\frac{\partial\sigma}{\partial t}+\frac{1}{m}\hspace{0.1cm}\frac{\partial S}{\partial x}\hspace{0.1cm}\frac{\partial\sigma}{\partial x}+
\frac{1}{2m}\left(\frac{\partial\sigma}{\partial x}\right)^{2}-\frac{i}{2m}\hspace{0.1cm}\frac{\partial^{2}\sigma}{\partial x^{2}}=
\frac{i}{2m}\hspace{0.1cm}\frac{\partial^{2}S}{\partial x^{2}}
\end{equation}   
with the boundary condition
\begin{equation}
\label{8}
\left(\frac{\partial\sigma}{\partial x}\right)_{x=0}=0
\end{equation}
Equations (\ref{5}) and (\ref{7}) are exact ones. The solution of the Hamilton-Jacobi equation (\ref{5}) can be found by 
conventional methods \cite{QUANT}
\begin{align}
\label{9}
S(x,t)=&-\frac{1}{2m}\int^{t}_{t_0}dt_{1}\left(p+{\cal E}_{0}t_{1}+\int^{t_1}_{0}dt_{2}{\cal E}(t_{2})\right)^{2}\nonumber\\
&+\left(p+{\cal E}_{0}t+\int^{t}_{0}dt_{1}{\cal E}(t_1)\right)x +(V-E)t_{0}-Vt
\end{align}
The functions $p(x,t)$ and $t_{0}(x,t)$ have to be defined from the conditions $\partial S/\partial p=0$ and 
$\partial S/\partial t_{0}=0$ what gives the following expressions
\begin{equation}
\label{10}
p(x,t)=i\sqrt{2m(V-E)}-{\cal E}_{0}t_{0}-\int^{t_0}_{0}dt_{2}{\cal E}(t_2)
\end{equation} 
and
\begin{equation}
\label{11}
mx=\int^{t}_{t_0}dt_{1}\left(p+{\cal E}_{0}t_{1}+\int^{t_1}_{0}dt_{2}{\cal E}(t_2)\right)
\end{equation}
Eq.~\ref{11} has to be inserted into Eqs.~\ref{10} and \ref{12}, what results in the final expression for the action
\begin{align}
\label{12}
S(x,t)&=-\frac{1}{2m}\int^{t}_{t_0}dt_{1}\left(i\sqrt{2m(V-E)}+(t_{1}-t_{0}){\cal E}_{0}
+\int^{t_1}_{t_0}dt_{2}{\cal E}(t_2)\right)^{2}\nonumber\\
&+x\left(i\sqrt{2m(V-E)}+(t-t_{0}){\cal E}_{0}
+\int^{t}_{t_0}dt_{1}{\cal E}(t_1)\right)+(V-E)t_{0}-Vt
\end{align}
where the function $t_{0}(x,t)$ is given by the equation
\begin{equation}
\label{13}
mx=i(t-t_{0})\sqrt{2m(V-E)}+(t-t_{0})^{2}\hspace{0.1cm}\frac{{\cal E}_{0}}{2}+\int^{t}_{t_0}dt_{1}(t-t_{1}){\cal E}(t_1)
\end{equation}
By means of the relation (a partial derivative is taken under the fixed lower index)
\begin{equation}
\label{14}
\left(\frac{\partial}{\partial t}\right)_{x}+
\frac{1}{m}\left(\frac{\partial S}{\partial x}\right)_{t}\left(\frac{\partial}{\partial x}\right)_{t}=
\left(\frac{\partial}{\partial t}\right)_{t_0}
\end{equation}
the equation (\ref{7}) for $\sigma$ in new variables $t_0$ and $t$ has the form
\begin{equation}
\label{15}
\left(\frac{\partial\sigma}{\partial t}\right)_{t_0}-\frac{1}{4(V-E)F^{2}}\left(\frac{\partial\sigma}{\partial t_{0}}\right)^{2}_{t}+
\frac{i}{4(V-E)F}\left(\frac{\partial}{\partial t_{0}}\hspace{0.1cm}\frac{1}{F}\hspace{0.1cm}
\frac{\partial\sigma}{\partial t_{0}}\right)_{t}=\frac{1+h(t_0)}{F\tau_{00}}
\end{equation}
Here new notations are introduced
\begin{equation}
\label{16}
F(t_{0},t)=1+i\frac{t_{0}-t}{\tau_{00}}\left(1+h(t_0)\right);\hspace{1cm}h(t)=\frac{{\cal E}(t)}{{\cal E}_{0}};\hspace{1cm}
\tau_{00}=\frac{\sqrt{2m(V-E)}}{{\cal E}_{0}}
\end{equation}    
In terms of new variables the boundary condition (\ref{8}) reads 
\begin{equation}
\label{17}
\left(\frac{\partial\sigma(t_{0},t)}{\partial t_{0}}\right)_{t_{0}=t}=0
\end{equation}
In semiclassical approximation $\sigma$ should be small comparing to a big classical action $S$ and it can be expanded into
a series
\begin{equation}
\label{18}
\sigma=\sigma_{1}+\sigma_{2}+...
\end{equation}
which is produced by Eq.~\ref{15}, where the last two terms in the left-hand side have to be considered as a perturbation. 
Now one can write
\begin{equation}
\label{19}
\sigma_{n}(t_{0},t)=\int^{t-t_{0}}_{0}d\eta\Phi_{n}(\eta,t_{0})+\int^{t_{0}}_{0}dt_{1}\Phi_{n}(0,t_{1})
\end{equation}
where
\begin{equation}
\label{20}
\Phi_{1}(t-t_{0},t_{0})=\frac{1+h(t_0)}{F(t_{0},t)\tau_{00}}
\end{equation}
and
\begin{equation}
\label{21}
\Phi_{2}(t-t_{0},t_{0})=\frac{1}{4(V-E)F^{2}}\left(\frac{\partial\sigma_{1}}{\partial t_{0}}\right)^{2}_{t}-
\frac{i}{4(V-E)F}\left(\frac{\partial}{\partial t_{0}}\hspace{0.1cm}\frac{1}{F}\hspace{0.1cm}
\frac{\partial\sigma_{1}}{\partial t_{0}}\right)_{t}
\end{equation}
From Eqs.~\ref{19} and \ref{20} one can obtain an explicit expression 
\begin{equation}
\label{22}
i\sigma_{1}(t_{0},t)=-\frac{1}{2}\ln F(t_{0},t)+\frac{i}{2\tau_{00}}\int^{t_{0}}_{0}dt_{1}\left(1+h(t_{1})\right)
\end{equation}
The expression for $\sigma_{2}$ can be easily obtained from Eqs.~\ref{19} and \ref{21} but it is too bulk and we do not write 
it here. The main parametric estimate at $n\geq 1$
\begin{equation}
\label{23}
i\sigma_{n}\sim\frac{1}{\left((V-E)\tau_{00}\right)^{n-1}}
\end{equation}
characterizes Eq.~\ref{18} as a typical semiclassical series since $(V-E)\tau_{00}\gg 1$. The pulse width is supposed to be 
order of $\tau_{00}$. Eqs.~\ref{12} and \ref{13} for the classical action and Eqs.~\ref{18} - \ref{22} for $\sigma$ enable to 
consider a decay of the metastable state under action of the non-stationary field ${\cal E}(t)$.
\section{CAUSALITY}
Suppose a pulse of the electric field has the form
\begin{equation}
\label{23b}
\tilde{{\cal E}}(t)={\cal E}(t)\Theta(t^{\prime}-t)+{\cal E}^{\prime}(t)\Theta(t-t^{\prime})
\end{equation}
The equation (\ref{23b}) can be considered in the complex time if to represent $\Theta$-functions in the form
\begin{equation}
\label{23c}
\Theta(t)=\left(\frac{1}{1+\exp(-\lambda t)}\right)_{\lambda\rightarrow +\infty}
\end{equation}
With the definition (\ref{23c}) the $\Theta$-function can be treated as $\Theta({\rm Re}t)$. As follows from here, the function 
$\tilde{{\cal E}}(t)$ at the complex $t$-plane depends on ${\cal E}(t)$ $({\cal E}^{\prime}(t))$ only to the left (right) of the 
vertical line ${\rm Re}\hspace{0.1cm}t=t^{\prime}$. If to chose the contours of integration in Eq.~\ref{12} 
{\it to the left of the vertical line with the real part $t$}, than the action $S(x,t)$ does not have an information how the 
non-stationary pulse behaves at moments later than $t$. This choice of contours of integration corresponds to the 
causality principle.           
\section{CLASSICAL ACTION}
\begin{figure}[p]
\begin{center}
\leavevmode
\epsfbox{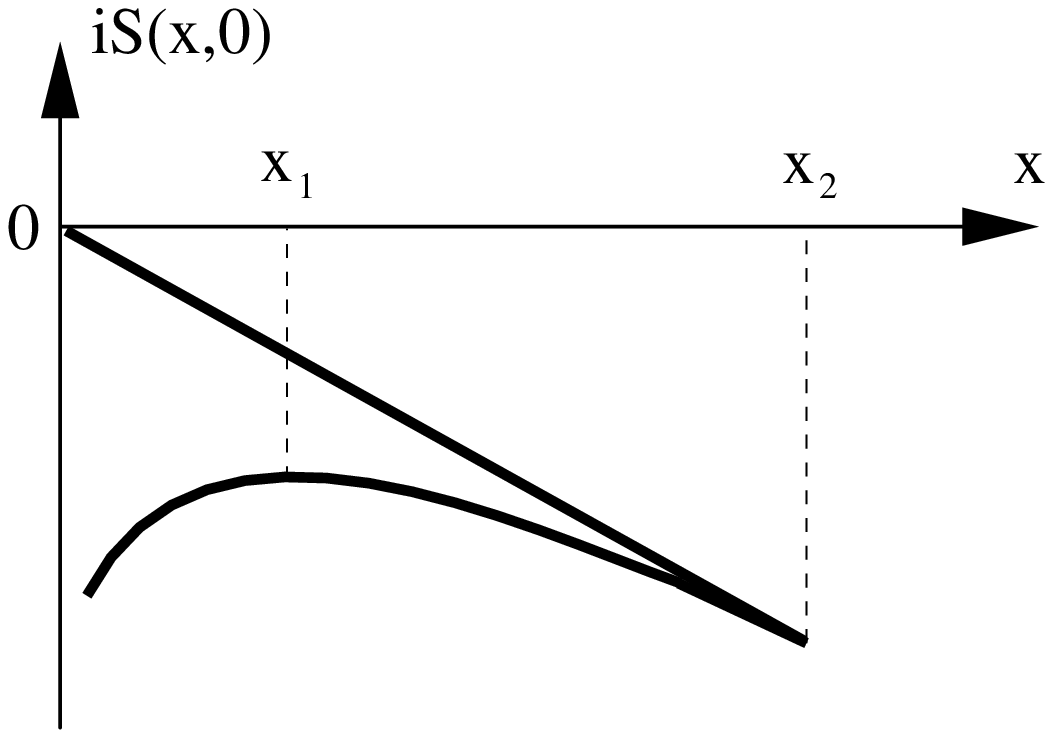}
\vspace{1cm}
\caption{A plot of the imaginary part of the action for the case of the
triangular potential barrier at $t=0$. The 
classical position of a particle before tunneling is $x=0$ and after 
tunneling is $x=x_{1}$ (a classical turning point).
In absence of a non-stationary pulse the two branches go over into
conventional increasing and decreasing WKB branches.}
\label{fig1}
\end{center}
\end{figure}

In this Section we consider only the classical action $S$ in Eq.~\ref{4a}. Under the signal (\ref{2}) the imaginary part of 
$S$ reaches its minimum value at some moment of time resulting in a maximum of the decay rate. For this reason we consider
first the action $S(x,0)$ at $t=0$. For a pulse symmetric in time $t_{0}(x,0)$ is an imaginary value and we introduce 
$\tau_{0}(x)=-it_{0}(x,0)$. At small amplitude of the signal (\ref{2}) essential values of $\tau_{0}$ are close to $\theta$ and 
the new variable
\begin{equation}
\label{23a}
z=1-\frac{\tau_{0}(x)}{\theta} 
\end{equation}
is convenient ($z\ll 1$). Below only integer values $n\geq 3$ in Eq.~\ref{2} and $\theta<\tau_{00}$ are considered. It follows 
from Eqs.~\ref{12} and \ref{13}
\begin{equation}
\label{24}
i\hspace{0.1cm}\frac{\partial S(x,0)}{\partial x}=-(\tau_{00}-\theta)\left(1-\left(\frac{z_1}{z}\right)^{n-1}\right){\cal E}_{0}
\hspace{0.1cm};\hspace{1cm}z_{1}=\left(\frac{{\cal E}\theta}{2^{n}(n-1)(\tau_{00}-\theta){\cal E}_{0}}\right)^{1/(n-1)}
\end{equation}
and
\begin{equation}
\label{25}
i\hspace{0.1cm}\frac{\partial^{2}S(x,0)}{\partial x^{2}}=\frac{\theta/(\tau_{00}-\theta)+(z_{2}/z)^{n}}{1-(z_{2}/z)^{n}}
\hspace{0.1cm}\frac{m}{\theta}\hspace{0,1cm};\hspace{1cm}z_{2}=\left(\frac{{\cal E}\theta}
{2^{n}(\tau_{00}-\theta){\cal E}_{0}}\right)^{1/n}
\end{equation}
In Eqs.~\ref{24} and \ref{25} the amplitude of the signal ${\cal E}$ is supposed to be small leading to small $z_1$ and $z_2$. 
As follows from Eq.~\ref{13}, the function 
$\tau_{0}(x)$ is determined by the relation
\begin{equation}
\label{26}
\frac{\partial mx}{\partial \tau_{0}}=(\tau_{00}-\theta){\cal E}_{0}-\frac{{\cal E}\theta}{2^{n}z^{n}}
\end{equation}
By means of Eqs.~\ref{24} and \ref{25} one can find the coordinate dependence of the action $S$, which is shown in 
Fig.~\ref{fig1}. 
Two branches in Fig.~\ref{fig1} 
in the limit ${\cal E}=0$ go over into conventional WKB wave functions
$\exp(\pm\int\mid p\mid dx)$. At the lower branch, where $\partial S/\partial x=0$, 
\begin{equation}
\label{27}
iS(x_{1},0)=(V-E)\hspace{0.1cm}\theta\left(1-\frac{\theta^{2}}{3\tau^{2}_{00}}\right)\hspace{0.1cm};\hspace{1cm}x_{1}=
\frac{{\cal E}_{0}\theta^{2}}{2m}
\end{equation}
At the common point, where $\partial^{2}S/\partial x^{2}\rightarrow\infty$,
\begin{equation}
\label{28}
iS(x_{2},0)=(V-E)\theta\left(1-\frac{\theta}{\tau_{00}}\right)^{2};\hspace{1cm}x_{2}=
\frac{{\cal E}_{0}\theta}{2m}\hspace{0.1cm}(2\tau_{00}-\theta)
\end{equation}
Near the common point $x_2$ the deviation of the action from the value (\ref{28}) is proportional to $(x_{2}-x)$ and in the 
second order to $\pm(x_{2}-x)^{3/2}$, what develops the two branches. 
\section{NON-SEMICLASSICAL CORRECTIONS}
For validity of the semiclassical approximation the inequalities
\begin{equation}
\label{28a}
\mid S\mid\gg\mid\sigma_{1}\mid\gg\mid\sigma_{2}\mid\gg ...
\end{equation}
should hold.  
Like in a static case, one can expect a violation of the semiclassical theory near the point $x=x_1$, where 
$\partial S/\partial x=0$, and the point $x=x_2$, where $\partial^{2}S/\partial x^{2}\rightarrow\infty$. Let us compare $\sigma$ with the 
classical action $S$ near these ``dangerous'' points. 

Under the condition $z_{1}\ll z\ll z^{n/(n+2)}_{2}$, as follows from Eqs.~\ref{20}-\ref{23}, 
\begin{equation}
\label{29}
i\sigma_{1}(x,0)=-\frac{1}{2}\ln\left(\left(1-\frac{\theta}{\tau_{00}}\right)\left(1-\left(\frac{z_{2}}{z}\right)^{n}\right)\right)-
\frac{\theta}{2\tau_{00}}
\end{equation}
and
\begin{equation}
\label{30}
i\sigma_{2}(x,0)=\frac{1}{48(V-E)\theta}\hspace{0.2cm}
\frac{3n(n+1)+n(2n-3)(z_{2}/z)^{n}}{z^{2}_{2}\hspace{0.1cm}(1-\theta/\tau_{00})\left(1-(z_{2}/z)^{n}\right)^{3}}
\left(\frac{z_{2}}{z}\right)^{n+2}
\end{equation}
An $x$-dependence in the right hand sides of Eqs.~\ref{29} and \ref{30} comes through $z$ according to Eq.~\ref{23a}. At the 
point $x=x_{1}$ $(z=z_{1})$ 
\begin{equation}
\label{31}
i\sigma_{1}(x_{1},0)=-\frac{1}{2}\ln\left(\frac{n-1}{z_{1}}\left(1-\frac{\theta}{\tau_{00}}\right)\right)-\frac{1}{2}-\frac{i\pi}{2}
\end{equation}
and
\begin{equation}
\label{32}
i\sigma_{2}(x_{1},0)=-\frac{1}{48(V-E)\theta z_{1}(1-\theta/\tau_{00})}\left(8(n-1)\left(1-\frac{\theta}{\tau_{00}}\right)^{2}+
\frac{n(2n-3)}{n-1}\right)
\end{equation}
A particle positioned in the well corresponds to $S(0,0)$ at the upper branch in Fig.\hspace{0.1cm}1 $(z\sim 1)$. A particle 
leaves the barrier under the condition $\partial S/\partial x=0$ at the point $x=x_{1}$ of the lower branch in Fig.\hspace{0.1cm}1 
$(z=z_{1})$. One should be sure the points $x=0$ and $x=x_{1}$ relate to the same semiclassical solution, on the other words,
it should be possible to find a way from $0$ to $x_{1}$ with no violation of the semiclassical conditions (\ref{28a}). 
Between real points $z\sim 1$ and $z=z_{1}$ there is only one ``dangerous'' point $z=z_{2}$, where, according to Eq.~\ref{30}, 
$\sigma\rightarrow\infty$ and the condition (\ref{28a}) breaks down. Nevertheless, the semiclassical approximation remains
valid if the condition (\ref{28a}) holds on some contour $\mid z-z_{2}\mid\sim z_{2}$ around the point $z_{2}$ in the complex 
$z$-plane. The point $z_{2}$ $(x=x_{2})$ is a branch point of the action, which has a contribution proportional to 
$(x_{2}-x)^{3/2}$, like a turning point in a static problem. The sequence of the Stokes and anti-Stokes lines \cite{HEADING} 
going from this point is qualitatively the same as in a static case. The condition (\ref{28a}) on the above circle can be 
written in the following approximate form if to put $z\sim z_{2}$ in Eq.~\ref{30} 
$(\theta<\tau_{00})$
\begin{equation}   
\label{33}
\left(\frac{\theta}{\tau_{00}-\theta}\right)^{n/2 -1}\frac{a_{n}}{((V-E)\theta)^{n/2}}\ll\frac{{\cal E}}{{\cal E}_{0}}\hspace{0.1cm};
\hspace{1cm}1\ll(V-E)\theta
\end{equation}
In the relation (\ref{33}) the numerical coefficient $a_{n}\sim 1$ at $n\sim 1$, but at big $n$ the coefficient $a_{n}$ 
increases, what prescribes to choose a not big $n$ for validity of the semiclassical approximation. The condition 
$\mid\sigma_{2}\mid\ll\mid\sigma_{1}\mid$ at the point $x=x_{1}$ is less rigorous. The semiclassical conditions (\ref{33}) require 
the pulse amplitude ${\cal E}$ to be not small. Remarkably, this amplitude can be still less than the static barrier field 
${\cal E}_{0}$. At lower ${\cal E}$, than one satisfying the relations (\ref{33}), one should expect the perturbation theory to 
be applicable.
\section{FINITE TIME}
Eq.~\ref{13} determines the function $t_{0}(x,t)$ and for the signal (\ref{2}) one can write at $x=x_{1}$ and $\mid t\mid<\theta$ 
\begin{equation}
\label{34}
t_{0}(x,t)=i\tau_{1}+\frac{z_{1}\theta^{2}}{2(\tau_{00}-\theta)}\left(i\hspace{0.1cm}\frac{t^{2}}{\theta^{2}}+
\frac{t^{3}}{\theta^{3}}\right)
\end{equation}
According to the causality principle, the contour of integration in Eq.~\ref{12} should be to the left of the time $t$ and
hence the condition ${\rm Re}\hspace{0.1cm}t_{0}<t$ results in the restriction $t>0$. At $t<0$ the semiclassical approach in its
present form is not valid since the integration penetrates ``in the future'' and this case requires further investigation.
The time dependence of the classical action can be found from Eq.~\ref{9} 
\begin{equation}
\label{35}
{\rm Im}\hspace{0.1cm}\frac{\partial S(x,t)}{\partial t}=-\frac{1}{2m}\hspace{0.1cm}{\rm Im}\left(i\sqrt{2m(V-E)}+(t-t_{0}){\cal E}_{0}+
\int^{t}_{t_{0}}dt_{1}{\cal E}(t_{1})\right)^{2}
\end{equation} 
At $x=x_{1}$ and $t=0$ the right hand side of Eq.~\ref{35} is zero. One can easily obtain
\begin{equation}
\label{36}
{\rm Im}\hspace{0.1cm}S(x_{1},t)={\rm Im}\hspace{0.1cm}S(x_{1},0)+(n-1)(V-E)\hspace{0.1cm}\frac{t^{4}}{\theta\tau^{2}_{00}}
\end{equation}
\section{THE TRANSITION PROBABILITY}
Suppose $w(t)$ is the probability to find a particle in the $\delta$-function potential and initially $w$ equals unity. The 
continuity equation reads $\partial w/\partial t=-(2/m){\rm Im}(\psi^{*}\partial\psi/\partial x)$, where the right hand side is 
taken at $x=x_{1}$. Using expression for the wave function
\begin{equation}
\label{37}
\psi(x,t)=\psi(0,t)\exp\left(iS(x,t)-iS(0,t)+i\sigma(x,t)-i\sigma(0,t)\right)
\end{equation} 
where
\begin{equation}
\label{38}
\psi(0,t)\simeq\left(2m(V-E)\right)^{1/4}\exp(-iEt)
\end{equation}
and by means of Eqs.~\ref{37} and \ref{38} one can obtain
\begin{equation}
\label{39}
\frac{\partial w(t)}{\partial t}=-\sqrt{\frac{8(V-E)}{m}}\left(\frac{\partial\hspace{0.1cm}{\rm Re}S}{\partial x}\exp\left(-2\hspace{0.1cm}
{\rm Im}\hspace{0.1cm}(S+\sigma)\right)\right)_{x=x_{1}}
\end{equation}
Eqs.~\ref{10} and \ref{12} give $\partial{\rm Re}S/\partial x={\cal E}_{0}t$ at $x=x_{1}$ and the main time dependence in the 
exponential of Eq.~\ref{39} follows from $S(x_{1},t)$ (Eq.~\ref{36}). Collecting the all terms, one can obtain finally
$(\theta<\tau_{00})$
\begin{align}
\label{40}
\frac{\partial w(t)}{\partial t}=-&\frac{2(V-E)t}{{\rm e}(n-1)^{n/(n-1)}(\tau_{00}-\theta)}\left(\frac{{\cal E}\theta}
{2(\tau_{00}-\theta){\cal E}_{0}}\right)^{1/(n-1)}\exp\left(-2(n-1)\frac{(V-E)t^{4}}{\theta\tau^{2}_{00}}\right)\nonumber\\
&\exp\left(-2(V-E)\theta\left(1-\frac{\theta^{2}}{3\tau^{2}_{00}}\right)\right)
\end{align}
As discussed in Section VI, Eq.~\ref{40} is valid only at $t>0$. The semiclassical conditions (\ref{28a}) are supposed to 
hold. According to Eq.~\ref{40}, the typical time scale of the output flux $\Delta t\sim(\theta\tau^{2}_{00}/(V-E))^{1/4}$ determines
the uncertainty of energy of outgoing particles $\Delta E\sim \Delta t^{-1}$ which is much smaller than the energy $E$. The 
decay rate $\partial w(t)/\partial t$ tends to its static value at $t\rightarrow\pm\infty$ having a maximum at some moment of time
\begin{equation}
\label{40a}
\left(\frac{\partial w}{\partial t}\right)_{max}\sim\exp\left(-2(V-E)\theta\left(1-\frac{\theta^{2}}{2\tau^{2}_{00}}\right)\right)
\end{equation}
The exponent in Eq.~\ref{40a}, according to its derivation, is a minimum value of the imaginary part of the classical 
action. A minimum value of action can be calculated, as known from classical mechanics, by means of trajectories 
satisfying Newton's equation. So, when we are not interested in the full time dependence of a decay rate but we need only 
its maximum value with an exponential accuracy, the method of classical trajectories can be used. This method is described
in the next Section. 
\section{METHOD OF COMPLEX TRAJECTORIES}
In this Section we consider penetration of incident particles through a potential barrier under action of a non-stationary
pulse. We restrict ourselves only by the main exponential approximation when one can use the semiclassical expression for 
a wave function $\psi(x,t)\sim\exp(iS(x,t))$. We consider here a particle flux on to the barrier shown in Fig.\hspace{0.1cm}2, but
the final result can be easily applied to decay of the metastable state through the triangular barrier (\ref{4}). The 
maximum value of the outgoing flux of particles can be calculated as a maximum with respect to time
\begin{equation}
\label{60}
W_{max}\sim{\rm max}\mid\exp(iS(x,t)-iS(x_{0},t))\mid^{2}
\end{equation}
Here $x$ is some coordinate to the right of the barrier, $x_{0}\rightarrow -\infty$, where ${\rm Im}S(x_{0},t)=0$, and in Eq.~\ref{60}
one can put $S(x_{0},t_{0})$ instead of $S(x_{0},t)$. The right hand side of Eq.~\ref{60} does not depend on $x$ and $x_{0}$. It 
is a function of $t$ only. Eq.~\ref{60}  corresponds to the extreme classical action, which can be found by method of 
classical trajectories $x(t)$ defined in the complex $t$-plane, since in real time there is no classical trajectory for an
under-barrier motion. The complex path $C$ is shown in Fig.\hspace{0.1cm}3. The real classical turning point is 
$x_{1}=x(t_{1})$, where $\partial x(t)/\partial t=0$. This point corresponds to the classical exit of a particle from under the 
barrier.  The real coordinate $x_{0}=x({\tilde t}_{0})$ is defined under the condition 
${\rm Re}\hspace{0.1cm}{\tilde t}_{0}=t_{0}\rightarrow -\infty$. The classical trajectory connects the points 
$\{x_{0},{\tilde t}_{0}\}$ and $\{x_{1},t_{1}\}$. The contour $C$ is symmetric with respect to the real axis. It is convenient to
write Eq.~\ref{60} in the form 
\begin{equation}
\label{61}
W_{max}\sim\mid\exp(iS(x_{1},t_{1})-iS(x_{0},t_{0}))\mid^{2}
\end{equation}
Since at $t\rightarrow -\infty$ the non-stationary field ${\cal E}(t)\rightarrow 0$, a connection between values of the action at
the points $t_{0}$ and ${\tilde t}_{0}$ is simple
\begin{equation}
\label{62}
S(x_{0},t_{0})=S(x_{0},{\tilde t}_{0})+({\tilde t}_{0}-t_{0})E
\end{equation}
where $E$ is the energy of an incident particle. According to Eqs.~\ref{61} and \ref{62}, the maximum amplitude value of the
outgoing flux of particles has now the form 
\begin{equation}
\label{63}
W_{max}\sim\exp(-A)
\end{equation}
where
\begin{equation}
\label{64}
A=-i\int_{C}dt\left(\frac{m}{2}\left(\frac{\partial x}{\partial t}\right)^{2}-V(x)+x{\cal E}(t)+E\right)
\end{equation}
is defined by means of trajectory satisfying Newton's equation $m\hspace{0.1cm}\partial^{2}x/\partial t^{2}+V^{\prime}(x)={\cal E}(t)$ in
the complex time. Due to symmetry of the contour $C$ the value of $A$ is real. The trajectory $x(t)$ should not be 
necessary real at all $t$, it should be real at least in vicinities of real points $x_{0}$ and $x_{1}$. On the left 
horizontal parts of the contour $C$, where ${\cal E}(t)=0$, $x(t)$ satisfies the equation 
\begin{equation}  
\label{65}
\frac{m}{2}\left(\frac{\partial x}{\partial t}\right)^{2}+V(x)=E
\end{equation}
and is expressed through the real functions $x=f(t-{\tilde t}_{0},E)$ (up) and $x=f(t-{\tilde t}^{*}_{0},E)$ (down), where $E$ is a 
real energy. Now one can formulate conditions how to choose the contour $C$: for given (at $t\rightarrow-\infty$) particle 
energy $E$ and the pulse shape ${\cal E}(t)$ one should find ${\rm Im}\hspace{0.1cm}{\tilde t}_{0}$ and the real turning point
$x_{1}=x(t_{1})$. Eq.~\ref{64} holds for a potential barrier $V(x)$, which is an analytical function of the variable $x$. Such
a barrier has no artificial restriction in coordinate (no singularity at a real $x$). For this reason, the equation 
(\ref{64}) can be interpreted as one accounting not only an under-barrier part but also some pre-barrier motion. 
\section{APPLICATION TO A TRIANGULAR BARRIER}
Eq.~\ref{64} is applicable to the case of a potential barrier $V(x)$, which is an analytical function of the variable $x$. 
In the case of the triangular barrier (\ref{4}), which is a non-analytical function, the all classical path of the particle
is restricted by an under-barrier motion. In this case the exponent $A$, instead of Eq.~\ref{64}, should be written in the 
form
\begin{equation}
\label{42}
A=2\hspace{0.1cm}{\rm Im}\int^{0}_{i\tau_{0}}dt\left(\frac{m}{2}\left(\frac{\partial x}{\partial t}\right)^{2}-V+x{\cal E}_{0}+
x{\cal E}(t)+E\right)
\end{equation}
where $x(t)$ is the trajectory satisfying the Newton equation in the complex time
\begin{equation}
\label{43}
m\frac{\partial^{2}x}{\partial t^{2}}-{\cal E}_{0}={\cal E}(t)
\end{equation}
The trajectory starts at the metastable well $x(i\tau_{0})=0$ with the boundary conditions 
\begin{equation}
\label{44}
\left(\frac{\partial x(t)}{\partial t}\right)_{i\tau_{0}}=i\sqrt{\frac{2(V-E)}{m}}
\end{equation}
For a symmetric pulse ${\cal E}(-t)={\cal E}(t)$ the velocity $\partial x/\partial t =0$ at $t=0$, at this point the particle 
escape from under the barrier, and this terminates the integration in Eq.~\ref{42}. The parameter $\tau_{0}$ has a meaning 
of under-barrier traversal time \cite{LANDAUER} and can be found from the equation 
\begin{equation}
\label{45} 
{\cal E}_{0}\tau_{0}+\int^{\tau_{0}}_{0}d\tau {\cal E}(i\tau)=\sqrt{2m(V-E)}
\end{equation}
Eq.~\ref{45} is equivalent to the condition $p=0$ following from Eq.~\ref{10}. Since the time is imaginary the function $A$ 
can be called the Eucledian action
\begin{equation}
\label{46}
A=2(V-E)\tau_{0}-\frac{{\cal E}^{2}_{0}}{3m}\tau^{3}_{0}-\frac{2{\cal E}_{0}}{m}\int^{\tau_{0}}_{0}\tau d\tau\int^{\tau}_{0}d\tau_{1}
{\cal E}(i\tau_{1})-\frac{1}{m}\int^{\tau_{0}}_{0}d\tau\left(\int^{\tau}_{0}d\tau_{1}{\cal E}(i\tau_{1})\right)^{2}
\end{equation}
The outgoing particle has the energy $E+\delta E$, where
\begin{equation}
\label{47}
\delta E=V-E-\left({\cal E}_{0}+{\cal E}(0)\right)\left(\frac{{\cal E}_{0}}{2m}\tau^{2}_{0}+\frac{1}{m}\int^{\tau_{0}}_{0}d\tau
\int^{\tau}_{0}d\tau_{1}{\cal E}(i\tau_{1})\right)
\end{equation}
After escape the barrier an action of the non-stationary field on the particle can be omitted since it is determined by 
the parameter ${\cal E}/{\cal E}_{0}$ which is much smaller than one governing the particle under the barrier and defined by 
the conditions (\ref{33}). In the absence of a non-stationary pulse the energy of outgoing particles has the same value 
$(\delta E=0)$ and the Eucledian action equals $A_{0}$ determined by the conventional WKB formula
\begin{equation}
\label{48}
A_{0}(E)=\frac{4}{3}\hspace{0.1cm}(V-E)\tau_{00}
\end{equation}
where $\tau_{00}$ is given by Eq.~\ref{16} and has a meaning of the under-barrier traversal time in the stationary case. The 
condition $\theta=\tau_{00}$, which is a particular case of Eq.~\ref{3}, sets some threshold energy 
\begin{equation}
\label{49}
E_{T}=V-\frac{\theta^{2}{\cal E}^{2}_{0}}{2m}
\end{equation}
As follows from Eqs.~\ref{46} and \ref{47}, the intensity and the energy of outgoing particles strongly depends on whether
the initial energy $E$ bigger ($\tau_{00}<\theta$) or smaller ($\theta<\tau_{00}$) than $E_{T}$. At $E_{T}<E$ the effect of 
the non-stationary signal on tunneling is weak and increases only in the vicinity of $E_{T}$
\begin{equation}
\label{50}
A=A_{0}\left(1-\frac{3{\cal E}}{(n-1)2^{n}{\cal E}_{0}}\hspace{0.1cm}\frac{1}{\left(1-\tau_{00}/\theta\right)^{n-2}}\right);
\hspace{0.7cm}\frac{\delta E}{V-E}=\frac{2{\cal E}}{(n-1)2^{n}{\cal E}_{0}}\hspace{0.1cm}
\frac{1}{\left(1-\tau_{00}/\theta\right)^{n-1}}
\end{equation}
At low energies $E<E_{T}$ the situation is very non-perturbative 
\begin{equation}
\label{51}
A=A_{0}(E_{T})+2(E_{T}-E)\theta;\hspace{1cm}\delta E=E_{T}-E
\end{equation}
what coincides with the exponent in Eq.~\ref{40a} obtained by a direct solution of the Hamilton-Jacobi equation. The 
energy dependence of the Eucledian action is shown in Fig.\hspace{0.1cm}4. This type of scenario of barrier penetration is 
shown schematically in Fig.\hspace{0.1cm}2 in the case of particle flux on the barrier. 
\section{SEPARATION OF QUANTA ABSORPTION AND TUNNELING}
For a monochromatic field ${\cal E}_{\omega}$ of frequency $\omega$ the total probability of penetration through a barrier can 
be approximately written as a product of two probabilities: absorption of $N$ quanta and tunneling (see the comment to
Eq.~\ref{1})
\begin{equation}
\label{52}
\frac{\partial w}{\partial t}\sim\left(\frac{{\cal E}_{\omega}}{{\cal E}_{0}\omega\tau_{00}}\right)^{2N}\exp\left(-A_{0}(E+\omega N)\right)=
\exp\left(-A(\omega,N)\right)
\end{equation}
For the pulse (\ref{1}) the amplitude ${\cal E}_{\omega}$ should be substituted by ${\cal E}(\omega\theta)^{n-1}\exp(-\omega\theta)$ and
the effective action becomes of the form
\begin{equation}
\label{53}
A(\omega,N)=A_{0}(E+\omega N)+2\ln\left(\frac{{\cal E}_{0}}{{\cal E}(\omega\theta)^{n-2}\exp(-\omega\theta)}\right)
\end{equation}
Semiclassical approximation corresponds to some optimum choice of $\omega$ in a continuous spectrum of the pulse and the 
number of quanta of this optimum frequency $N$, which provide a minimum of $A(\omega, N)$. The condition 
$\partial A(\omega,N)/\partial N =0$ gives the following relations
\begin{align}
\label{54}
&\delta E=\omega N=(E_{T}-E) -\frac{2\theta(V-E)}{\omega\tau_{00}}
\ln\left(\frac{{\cal E}_{0}}{{\cal E}(\omega\theta)^{n-2}}\right)\nonumber\\
&A=A_{0}(E_{T})+2(E_{T}-E)\theta-\frac{2(V-E)}{\omega}\left(1-\frac{\theta^{2}}{\tau^{2}_{00}}\right)
\ln\left(\frac{{\cal E}_{0}}{{\cal E}(\omega\theta)^{n-2}}\right)
\end{align}
The further minimization, with respect to $\omega$, give an infinite (in this approach) value of $\omega$ indicating the 
logarithmic terms in Eqs.~\ref{54} to be small and hence Eqs.~\ref{54} coincide with the result (\ref{51}). 

Let us consider another example, when such simple approach also give a correct (with an exponential accuracy) decay rate.
Suppose the Gaussian pulse
\begin{equation}
\label{55}
{\cal E}(t)={\cal E}\exp(-\Omega^{2}t^{2})
\end{equation}
acts on a particle in the stable potential well (\ref{4}) with ${\cal E}_{0}=0$. Than in the effective action 
\begin{equation}
\label{56}
A(\omega,N)=2N\ln\frac{\omega\sqrt{m(V-E)}}{{\cal E}_{\omega}}
\end{equation}
one should put ${\cal E}_{\omega}\rightarrow {\cal E}\exp(-\omega^{2}/\Omega^{2})$, according to the Fourier harmonic of the pulse, and
$N=(V-E)/\omega$, since in this case there is no tunneling and a particle should reach the top of the barrier. This leads to 
the relation
\begin{equation}
\label{57}
A\left(\omega,\frac{V-E}{\omega}\right)=2(V-E)\left(\frac{\omega}{4\Omega^{2}}+\frac{1}{\omega}\ln\frac{\omega\sqrt{m(V-E)}}{{\cal E}}\right)
\end{equation}
The minimization of this expression with respect to $\omega$ gives the optimum value of $A$
\begin{equation}
\label{58}
A=\frac{2(V-E)}{\Omega}\left(\ln\frac{\Omega\sqrt{m(V-E)}}{{\cal E}}\right)^{1/2}
\end{equation} 
the optimum pulse frequency, and the optimum number of absorbed quanta
\begin{equation}
\label{59}
\omega_{opt}=2\Omega\left(\ln\frac{\Omega\sqrt{m(V-E)}}{{\cal E}}\right)^{1/2};\hspace{1cm}N_{opt}=\frac{V-E}{\omega_{opt}}
\end{equation}
The result (\ref{58}) coincides with calculation of the Eucledian action for the pulse (\ref{55}) by semiclassical methods 
developed above (calculations are not put in this paper). One can see from here, the decay rate under action of a 
non-stationary pulse can be calculated with an exponential accuracy on the base of simple arguments of optimum frequency 
and number of quanta. This approach of separation of quanta absorption and subsequent tunneling, described in this Section,
works only for a potential $V(x)$ which is {\it not an analytical function} of the variable $x$ like the potential (\ref{4}). In 
this case one can use an interpretation of quanta absorption at some point $x$ (position of singularity of $V(x)$ on the 
real axis). When $V(x)$ is an analytical function there is no such particular point, the situation is more complicated, and
the method of simple separation of absorption and tunneling does not work, since the quantum interference of these 
processes becomes very non-trivial. The case of analytical potential is considered in the next Section.
\section{WEAK NON-STATIONARY SIGNAL}
Let us go back to an analytical potential barrier $V(x)$. When ${\cal E}(t)=0$ the contour $C$ is reduced to the contour 
$C_{0}$ shown in Fig.\hspace{0.1cm}3, which consists of the vertical part between the points 
$\pm i\hspace{0.1cm}{\rm Im}\hspace{0.1cm}{\tilde t}_{0}$ and the horizontal semi-infinite lines at ${\rm Re}t<0$. In this static case
\begin{equation}
\label{66}
{\rm Im}\hspace{0.1cm}{\tilde t}_{0}=\sqrt{\frac{m}{2}}\int\frac{dx}{\sqrt{V(x)-E}}
\end{equation}
where the integration goes between two classical turning points determined by the relation $V(x)=E$. Eq.~\ref{64} at
${\cal E}(t)=0$ determines the conventional WKB exponent by means of the unperturbed Lagrangian $L_{0}$
\begin{align}
\label{67}
&A_{0}=-i\int_{C}dtL_{0}=2\sqrt{2m}\int dx\sqrt{V(x)-E}\nonumber\\
&L_{0}=\frac{m}{2}\left(\frac{\partial x_{0}}{\partial t}\right)^{2}-V(x_{0})+E
\end{align}
Here $x_{0}(t)$ is the classical trajectory determined at all $t$ by Eq.~\ref{65}. The small pulse ${\cal E}(t)$ results in the
perturbed trajectory $x_{0}(t)+\delta x(t)$. The perturbation in Eq.~\ref{64} has the form
\begin{equation}
\label{68}
A=-i\int_{C}L_{0}+\int_{C}x_{0}(t){\cal E}(t)+
m\left(\frac{\partial x_{0}}{\partial t}\delta x\right)({\tilde t}^{*}_{0})-
m\left(\frac{\partial x_{0}}{\partial t}\delta x\right)({\tilde t}_{0})
\end{equation}
The velocities $\partial x/\partial t$ at ${\tilde t}^{*}_{0}$ and ${\tilde t}_{0}$ are real and
\begin{equation}
\label{69}
{\rm Im}\hspace{0.1cm}\delta x({\tilde t}_{0})=-\left(\frac{\partial x_{0}}{\partial t}\delta{\tilde t}_{0}\right)({\tilde t}_{0})
\end{equation}
Here $\delta{\tilde t}_{0}$ is a variation of ${\tilde t}_{0}$ due to the pulse given by Eq.~\ref{66}. One can easily see that
\begin{equation}
\label{70}
\int_{C}dtL_{0}=\int_{C_{0}}dtL_{0}-2i\left(\frac{\partial x_{0}}{\partial t}\right)^{2}({\tilde t}_{0})\hspace{0.1cm}{\rm Im}
\hspace{0.1cm}\delta{\tilde t}_{0}
\end{equation}
Collecting Eqs.~\ref{68}-\ref{70}, one can obtain
\begin{equation}
\label{71}
A=A_{0}+\delta A\hspace{0.1cm};\hspace{1cm}\delta A= -i\int_{C}dt\hspace{0.1cm}{\cal E}(t)x_{0}(t+\Delta t)
\end{equation}
Here we keep the argument shift $\Delta t$ of the unperturbed solution, satisfying Eq.~\ref{65}, determined in the way the 
classical turning point $t_{1}=-\Delta t$. The method of classical trajectories produces a minimum value of $A$, this means 
the shift $\Delta t$ to be found from the minimization condition
\begin{equation}
\label{72}
\frac{\partial \delta A}{\partial \Delta t}=0
\end{equation}
A meaning of the minimization condition (\ref{72}) can be clarified in the following way. According to classical mechanics, 
the variation of the particle energy is $\partial E/\partial t={\cal E}(t)\partial x_{0}/\partial t$. At $t={\tilde t}_{0}$ and at 
$t={\tilde t}^{*}_{0}$, when the non-stationary field is zero, energy should have the same values, that is
\begin{equation} 
\label{73}
\int_{C}dt\hspace{0.1cm}{\cal E}(t)\frac{\partial x_{0}(t+\Delta t)}{\partial t}=0
\end{equation}
Eq.~\ref{73} coincides with Eq.~\ref{72}. If the minimization condition (\ref{72}) violates the particle energy $E$ would
acquire an imaginary part. In summary of this Section, in case of small non-stationary field ${\cal E}(t)$ one can use the 
perturbation approach (\ref{71}) with the further minimization (\ref{73}).
\section{ANALYTICAL PROPERTIES OF TRAJECTORIES}
We consider the potential barrier
\begin{equation}
\label{74}
V(x)=\frac{V}{\cosh^{2} x/a}
\end{equation}
The classical unperturbed trajectory satisfies the relations \cite{MELN,MELN1}
\begin{equation}
\label{75}
\frac{\partial x_{0}(t+\Delta t)}{\partial t}=\frac{a\omega\sinh\omega(t+\Delta t)}{\sqrt{\cosh^{2}\omega(t+\Delta t)+E/(V-E)}}
\hspace{0.1cm};\hspace{1cm}\omega^{2}=\frac{2E}{ma^{2}}
\end{equation}
and is an analytical function of the complex variable $t$ having the branch points at $t=t_{s},t^{*}_{s}$ where
\begin{equation}
\label{76}
t_{s}=i\tau_{s}-\frac{1}{\omega}\ln\frac{\sqrt V +\sqrt E}{\sqrt{V-E}}\hspace{0.1cm}-\Delta t;\hspace{1cm}\tau_{s}=\frac{\pi}{2\omega}
\end{equation}
Close to the branch point $t_{s}$ the trajectory has the form
\begin{equation}
\label{77}
x_{0}(t+\Delta t)=-\frac{i\pi a}{2}+a\sqrt{2\omega(t_{s}-t)\sqrt{V/E}}
\end{equation}
The cut is shown in Fig.\hspace{0.1cm}3 by the dashed horizontal line. Now the integral, defining $\delta A$ in Eq.~\ref{71}, 
can be calculated on base of analytical properties.
\section{TUNNELING PROBABILITY}
Let us choose the non-stationary pulse in the form (\ref{2}) with $n=2$ and the potential barrier (\ref{74}). Then the 
integrand in Eq.~\ref{71} has 
singularities of two types in the complex $t$-plane: $t_{s}$ comes from the analytical function $x_{0}(t+\Delta t)$ and 
$i\theta$ comes from the analytical function ${\cal E}(t)$. There are different positions of the contour $C$ with respect
to those singular points giving rise to different branches of the extreme action. The branch, giving the minimum value of 
$A$, results from the position shown in Fig.\hspace{0.1cm}3, when the contour $C$ goes between the two singularities. This 
very branch determines the effect. We consider here only the case ${\rm Im}t_{s}<\theta$. The contour $C$ can be deformed up 
and in the limit $(\theta-\tau_{s})\ll\theta$ the pole circle at $t=i\theta$ gives the main contribution
\begin{equation}
\label{78}
\delta A=\pi{\cal E}a\tau^{2}_{s}\left(\frac{V}{E}\right)^{1/4}{\rm Re}\hspace{0.1cm}\sqrt{\frac{\omega}{2(t_{s}-i\theta)}}
\end{equation} 
According to Eq.~\ref{76}, $t_{s}$ depends on $\Delta t$ and the minimization (\ref{72}) produces
\begin{equation} 
\label{79}
\Delta t=-\frac{\theta-\tau_{s}}{\sqrt 3}
\end{equation}
The moment $t_{1}=-\Delta t$ is the delay time of outgoing particles from under the barrier. The correction $\delta A$ in 
Eq.~\ref{71} has the form
\begin{equation}
\label{80}
\delta A =-\frac{\pi}{4}{\cal E}a\tau^{2}_{s}\left(\frac{3V}{E}\right)^{1/4}\sqrt{\frac{3\omega}{\theta-\tau_{s}}}
\end{equation}
Finally, Eqs.~\ref{63}, \ref{67}, \ref{71}, and \ref{80} determine the maximum outgoing flux of particles tunneling through 
the potential barrier (\ref{74}) in the case of a weak non-stationary signal (\ref{2}) with $n=2$. Eq.~\ref{80} is analogous 
to the formula (\ref{50}) for the triangular potential. When $\theta\rightarrow{\rm Im}\hspace{0.1cm}t_{s}$ the perturbation theory 
breaks down and the result becomes to be very non-linear function of the non-stationary pulse like Eq.~\ref{51}. We see 
that the relation (\ref{3}) plays a crucial role in physics of tunneling under non-stationary conditions. For the potential
barrier (\ref{74}) $t_{s}$ appears in a ``natural'' way as a result of analytical properties, whereas for a non-analytical 
potential it is determined by the time of motion between a turning point and a point of the non-analytical singularity of 
the potential ($\tau_{00}$ in the case of the triangular potential (\ref{4})).
\section{DISCUSSIONS AND CONCLUSIONS}
In the case of the triangular potential barrier the semiclassical theory in a non-stationary case is constructed based on
classical trajectories in the complex time. This becomes possible since for such barrier non-semiclassical corrections
can be calculated exactly. The conditions of applicability of the semiclassical approach (softly varying and not very small
pulse) are not surprising. Much less trivial matter is the way of construction of the trajectory method. The condition
(\ref{3}) of coincidence of singularity of a trajectory and of a field leads to unexpected conclusions. The threshold energy
$E_{T}$, set by the condition (\ref{3}), divides the all incident flux of particles by two groups: (i) particles with 
$E>E_{T}$ passes the barrier practically like in a stationary case, (ii) particles with lower energies $E<E_{T}$ are affected 
strongly by the pulse which collects all particles after passing the barrier at the same energy $E_{T}$ regardless their 
initial energy. This enables to use short pulses, adapted to the necessary energy level by the relation (\ref{3}), to get 
outgoing particles collected at this level at some moment of time. In other words, adapted signals can be used for a 
selective control of tunneling: in solids and molecules it excites only some particular bonds leaving other bonds 
non-excited. There is an obvious advantage of use soft signals for control of tunneling. The hard pulse of the type 
(\ref{2}) cannot control tunneling selectively, since it kicks up all particles to the top of the barrier. When the pulse 
(\ref{2}) serves as an envelope for the monochromatic signal of the frequency $\Omega\sim 1/V$ it provides in a non-selective 
manner equal $\Omega$-shifts of all escaping particles with respect to their incident energies. It is remarkable, the soft 
Gaussian signal ${\cal E}(t)={\cal E}\exp(-t^{2}/\theta^{2})$ cannot produce a strong enhancement of tunneling when its amplitude 
is much less than a static field of a potential barrier. This is an indication of importance of the analytical structure 
of the signal (\ref{2}). This can be understood in another way looking at Eq.~\ref{1}, where the Fourier harmonics of the 
Gaussian signal ${\cal E}_{\Omega}\sim{\cal E}\exp(-\Omega^{2}\theta^{2}/4)$ do not result in divergence in $\Omega$.

The static electric field in solids and molecules can be estimated as ${\cal E}_{0}\sim 10^{7}{\rm V/cm}$, a typical pulse width 
is in the range of tens of femtoseconds, and the amplitude of the electric field of the pulse can be chosen as 
${\cal E}\sim 10^{4}-10^{5}{\rm V/cm}$, what is reachable in experiments.

The developed theory is also applicable to quantum mechanical over-barrier reflection of particles. The reflection of 
classical waves (electromagnetic, hydrodynamic, etc.) from a spatially-smooth medium also may be described by the above 
theory, when the medium is influenced by an adapted signal.
\section{ACKNOWLEDGMENT} 
I am grateful to S. Obukhov, J. Krause, A. Efimov, and E. Ryabov for stimulating discussions

\newpage

\newpage
\begin{figure}[p]
\begin{center}
\leavevmode
\epsfbox{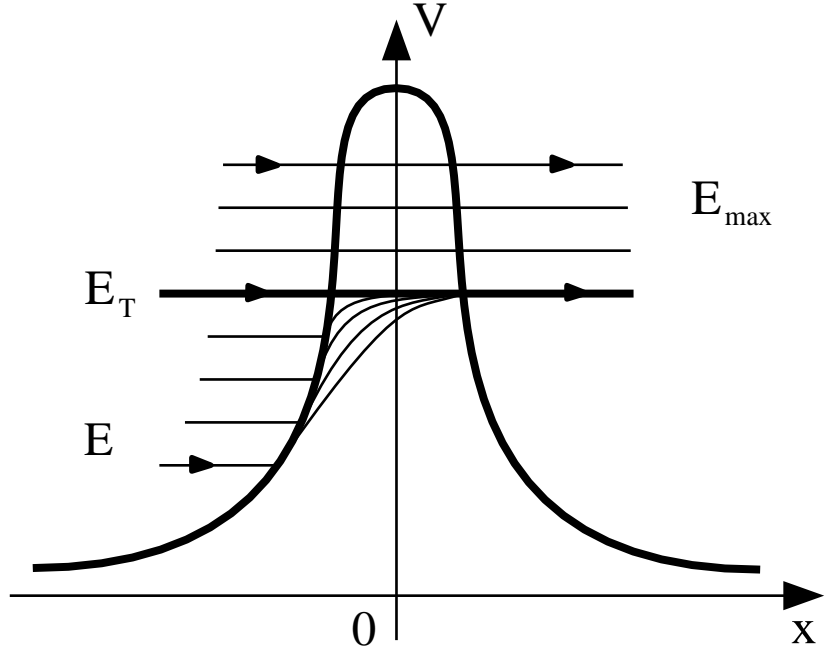}
\vspace{1cm}
\caption{A snap-shot of a particle motion through the potential barrier under the action of a non-stationary pulse at a
moment of the maximum tunneling flux. $E_{max}$ is the maximum (in time) energy of escaped particles. All particles with 
the incident energies $E<E_{T}$ are collected at the threshold level $E_{T}$ after passing the barrier. A motion of 
particles with $E>E_{T}$ is violated a little.}
\label{fig2}
\end{center}
\end{figure}

\newpage
\begin{figure}[p]
\begin{center}
\leavevmode
\epsfbox{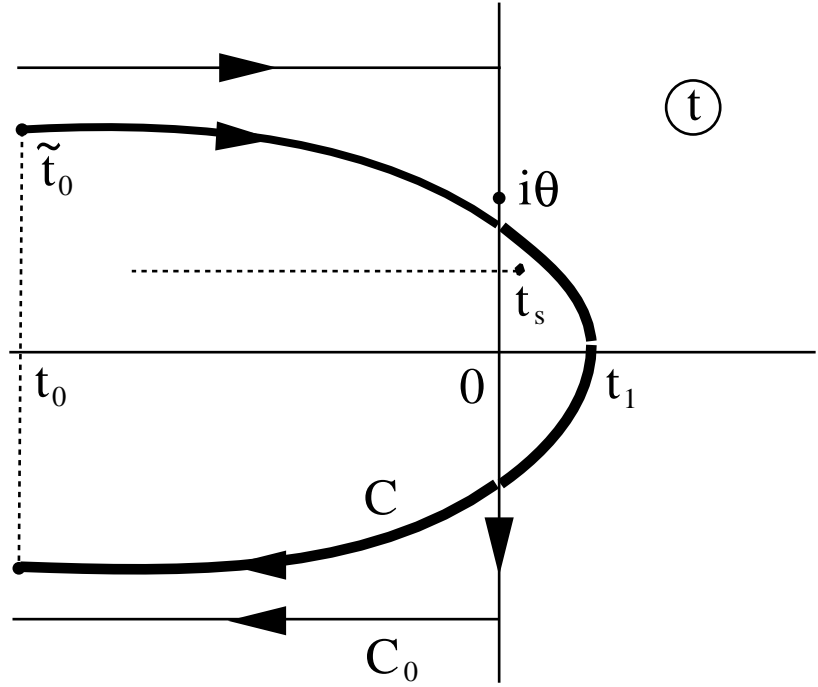}
\vspace{1cm}
\caption{The contours of integration $C$ and $C_{0}$ (in absence of a pulse) are shown in the plane of complex time. 
$i\theta$ is the position of singularity of the non-stationary pulse and $t_{s}$ is the branch point singularity of the 
classical trajectory. The cut is denoted by the dashed horizontal line.}
\label{fig3}
\end{center}
\end{figure}

\newpage
\begin{figure}[p]
\begin{center}
\leavevmode
\epsfbox{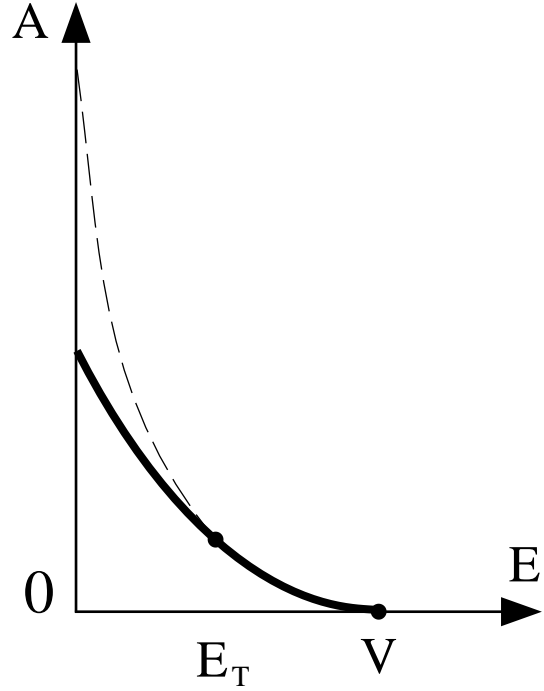}
\vspace{1cm}
\caption{The energy dependence of the exponent $A(E)$ (solid line) which determines the maximum (in time) of the tunneling
probability $W_{max}\sim\exp(-A(E))$. The dashed line is the plot of the exponent $A_{0}(E)$ in the absence of a pulse. It 
merges the solid line at $E>E_{T}$.}
\label{fig4}
\end{center}
\end{figure}

\end{document}